\documentclass[aps,prb,reprint,showpacs,showkeys,floatfix]{revtex4-1}

\usepackage{graphicx}
\usepackage{dcolumn}
\usepackage[latin1]{inputenc}
\usepackage{amsmath}
\usepackage{amsfonts}
\usepackage{upgreek}
\usepackage{color}
\usepackage{subfigure}

\begin{document}

\title{Prediction of confined and controllable Bloch points in nanocubes of chiral magnets}
\author{Michalis Charilaou}

\email{michalis.charilaou@louisiana.edu}
\affiliation{Department of Physics, University of Louisiana at Lafayette, Lafayette, LA 70504}

\begin{abstract}
This work predicts that individual Bloch points can be created and stabilized by magnetostatic and chiral interactions in nanocuboids, confined in between two chiral bobbers of opposing polarity. The Bloch point can be moved by an external magnetic field of moderate strength but only if the field strength is enough to overcome a pinning potential that results from intrinsic exchange forces and extrinsic surface effects. The Bloch point can be driven by the external field reversibly, in a direction opposing the field, and it remains stable up to moderate field strengths. At a critical field strength the Bloch point escapes through one of the surfaces, leaving behind a collinear magnetization configuration, and upon removing the field a new Bloch point is formed. These findings highlight the topological diversity in nanostructures and show that a Bloch point, despite its zero-dimensionality, couples to external fields via a substantial magnetic volume around it. The control of topological point defects has technological implications with regards to reversibly movable nanomagnetic textures and their associated emergent electrodynamics.
\end{abstract}

\maketitle

\section{Introduction}
The past few decades have seen a cascade of discoveries that shed new light on the role of topology in condensed matter \cite{wang2017}, particularly with regards to magnetism \cite{braun2012,ackerman2017,zang2018} and the identification of surprisingly stable magnetization textures in the form of quasiparticles, such as skyrmions (Sk) \cite{roessler2006,hellman2017,fert2017,hoffmann2017,everschor2018,koshibae2016} and chiral bobbers (ChB) \cite{kiselev2015,kiselev2018}. In addition to the fundamental insight provided by the pursuit of topologically non-trivial quasiparticles, the stability and mobility of these nanomagnetic objects makes them highly promising candidates as non-volatile components in novel spintronics applications \cite{pacheco2017,kiselev2018,sampaio2013,fert2013,moutafis2016}. Before any sustainable technology can be developed, however, a deep understanding of the physics of topologically non-trivial nanomagnetic objects is essential.

\begin{figure}
	\centering
\includegraphics[width=0.9\columnwidth]{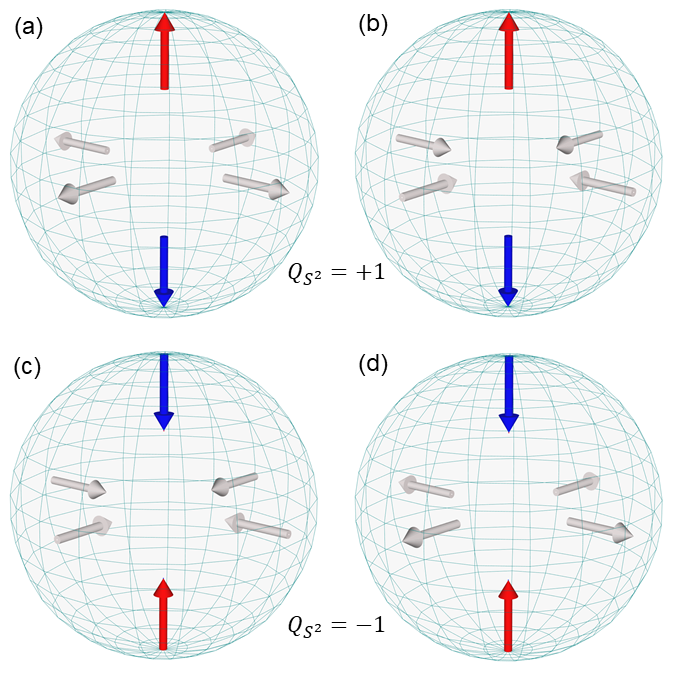} 
\caption{Schematic of the configuration of the local magnetization around radial Bloch points with a winding number, or topological charge, of $+1$ (a and b) and $-1$ (c and d), calculated with equation (4). }
\label{BPoints}
\end{figure}

The one most puzzling topologically non-trivial object that has been debated in the research community for more than 50 years \cite{feldtkeller1965a,feldtkeller1965b,doering1968,polyakov1974,arrott1979,malozemoff1979,volovik1987,kotiuga1989} is the singularity of the Bloch point \cite{milde2013,heyderman2017,kanazawa2017,tokura2019}, a topological point-defect where the magnetization vanishes and the micromagnetic continuum theory \cite{brown1963} breaks down. Bloch points occur in different topologically-equivalent configurations, e.g. as radial hedgehogs (see Fig. \ref{BPoints}) or as vortex hedgehogs \cite{zang2018}, and their creation is also associated to the mediation of topological phase transitions, e.g. breaking of skyrmion lines \cite{milde2013,schuette2014} or reversal of vortex states \cite{thiaville2003,hertel2007}, in a wide range of materials and nanostructures \cite{braun1999,dacol2014,hertel2016,charilaou2017,charilaou2018,im2019,beg2019}. 

Recent experiments have witnessed BPs in materials through a variety of techniques involving transport measurements \cite{kanazawa2016}, neutrons \cite{kanazawa2017,tokura2019}, and x-rays \cite{heyderman2017,im2019}, but despite the recent progress on modeling, imaging and detecting BPs, an effective way of creating and controlling individual Bloch points remains challenging. Importantly, the mechanism by which a well defined state, e.g. collinear ferromagnetic, can be wound in a way that creates a Bloch point t is still poorly understood.

This work predicts, by means of micromagnetic and atomistic simulations, that the collinear ferromagnetic state in nanocubes of appropriate dimensionality is modified by magnetostatics and chiral interactions in a way that results in a pair of chiral bobbers with opposite polarity, where the meeting point of the two ChBs corresponds to a Bloch point. The BP is confined inside the cuboid and can be moved up and down by an external magnetic field, whereas upon removal of the external field the BP returns to its equilibrium position that is determined by a pinning potential due to intrinsic interactions and surface effects. The simulations suggest that the only two factors that need to be controlled for the confinement of the BP is the size of the cube, which should fit one complete period of the magnetization texture, and the initial state.

\section{Computational Model}\label{methods}
For the simulations, the chiral magnet FeGe was chosen because it has been studied extensively \cite{everschor2018} and has a high Curie temperature \cite{wilhelm2011}, making it an attractive system both for basic research and for potential applications. The total micromagnetic energy density contains contributions from the ferromagnetic exchange with stiffness $A$, Dzyaloshinskii-Moriya interaction (DMi) with strength $D$, coupling to an external magnetic field $\mathbf{H}_\mathrm{ex}$, and magnetostatic dipole-dipole interactions via a local demagnetizing field $\mathbf{H}_\mathrm{d}$, 
\begin{align}\label{micromag_en}
\mathcal{E}=&A \left(\nabla \mathbf{m}\right)^2+D \mathbf{m} \cdot \left( \nabla \times \mathbf{m} \right)- \mu_0M_\mathrm{s}\mathbf{H}_\mathrm{ex}\cdot \mathbf{m}\nonumber \\ &-\frac{1}{2} \mu_0M_\mathrm{s}  \mathbf{H}_\mathrm{d} \cdot \mathbf{m} \; ,
\end{align}
where $M_\mathrm{s}$ is the saturation magnetization and $\mathbf{m}= {\bf M}/M_s$ is the unit vector of the magnetization parametrized by the polar and azimuthal angles $\theta$ and $\phi$, respectively, as ${\bf m} = ( \sin \theta \cos \phi, \sin \theta \sin \phi, \cos \theta)$. The system was discretized in a finite-difference mesh and the magnetization dynamics were computed using the software package Mumax3 \cite{mumax3} by numerically integrating the Landau-Lifshitz-Gilbert (LLG) equation of motion
\begin{equation}
\partial_t  \mathbf{m}= -\gamma \, \mathbf{m} \times \mathbf{H}_\mathrm{eff} +\alpha \, \mathbf{m} \times \partial_t \mathbf{m} \; ,
\end{equation}
where $\alpha$ is the dimensionless damping parameter (here set to 0.1), $\gamma$ is the electron gyromagnetic ratio, and $ \mu_0\mathbf{H}_\mathrm{eff}=-\partial_\mathbf{m} \mathcal{E}/M_\mathrm{s}$ is the effective field in the system, consisting of both internal and external contributions. The material parameters for FeGe were taken from the literature \cite{ericsson1981,yamada2003,wilhelm2011,beg2015}: $A=8.78$~pJ/m, $M_\mathrm{S}=385$~kA/m, and $D=1.58$~mJ/m$^2$. The ratio between $A$ and $D$ determines the helical pitch length $\lambda=4\pi A/\left|D\right|$, i.e. the periodicity of the spin texture, which for FeGe is $\sim 70$ nm. Hence, the cuboids discussed in this paper had a cross-sectional area of 70 nm $\times$ 70 nm, i.e., able to fit exactly one period of the spin texture. For high-resolution simulations, a cell size of 1 nm was chosen, much smaller than the exchange length $\delta_\mathrm{ex}=\sqrt{2A/\mu_0 M_\mathrm{s}^2}\approx 10$ nm, with occasional checks with other cell sizes to confirm numerical stability. 

\section{Results and Discussion}
\begin{figure}
	\centering
\includegraphics[width=1.0\columnwidth]{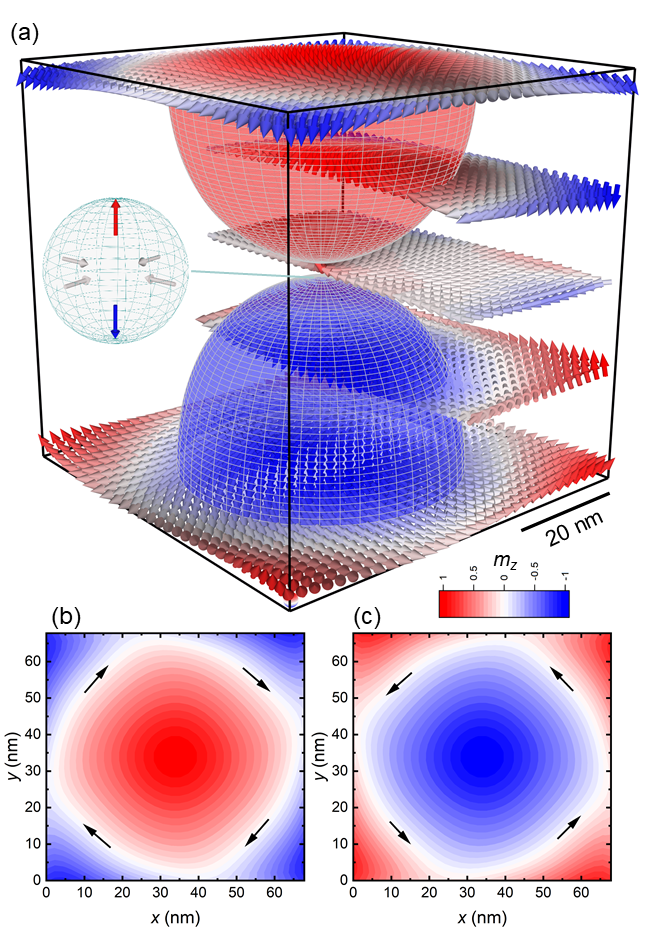} 
\caption{(a) Vector plot of the confined Bloch point in a 70 nm wide FeGe cube showing how the (b) top and (c) bottom surfaces contain a skyrmionic texture of the same handedness but opposite polarity, which extend below the surfaces in the form of chiral bobbers that intersect at the Bloch point. The inset to (a) shows the local configuration around the BP. In panel (a) only selected layers are shown for visual clarity.}
\label{MM_Main}
\end{figure}
In the simulations, the magnetization is initialized in the uniform ferromagnetic (FM) state, with the moments perpendicular to one of the cube's faces, and the LLG equation is integrated for several nanoseconds to find the equilibrium state. Note that the $z$ axis is parallel to the initial direction of the FM moment. The FM configuration is a high-energy state, therefore both DMi and dipolar interactions tilt the magnetic moments towards the faces, i.e., in the $xy$ plane, to lower the energy. The moments on the top surface curl in a left-handed sense and form a vortex-like texture, while the moments on the bottom surface curl in a right-handed sense. The opposite handedness of the textures on the two surfaces minimizes the magnetostatic term but frustrates the DMi, the sign of which favors left-handed chirality. Hence, while the DMi term is minimized with the left-handed winding of the top surface and further curls the magnetic moments to form a skyrmionic texture, the right-handedness at the bottom surface costs an energy penalty of $2D$ and therefore the polarity of the moments in the center of the surface is reversed to restore the DMi-favored handedness and form a left-handed skyrmionic texture (the same handedness as that of the top surface) with opposite polarity (see Fig. \ref{MM_Main}). In this way, magnetostatics and DMi, as well as the ferromagnetic exchange, are satisfied. 
\begin{figure}
	\centering
\includegraphics[width=1.0\columnwidth]{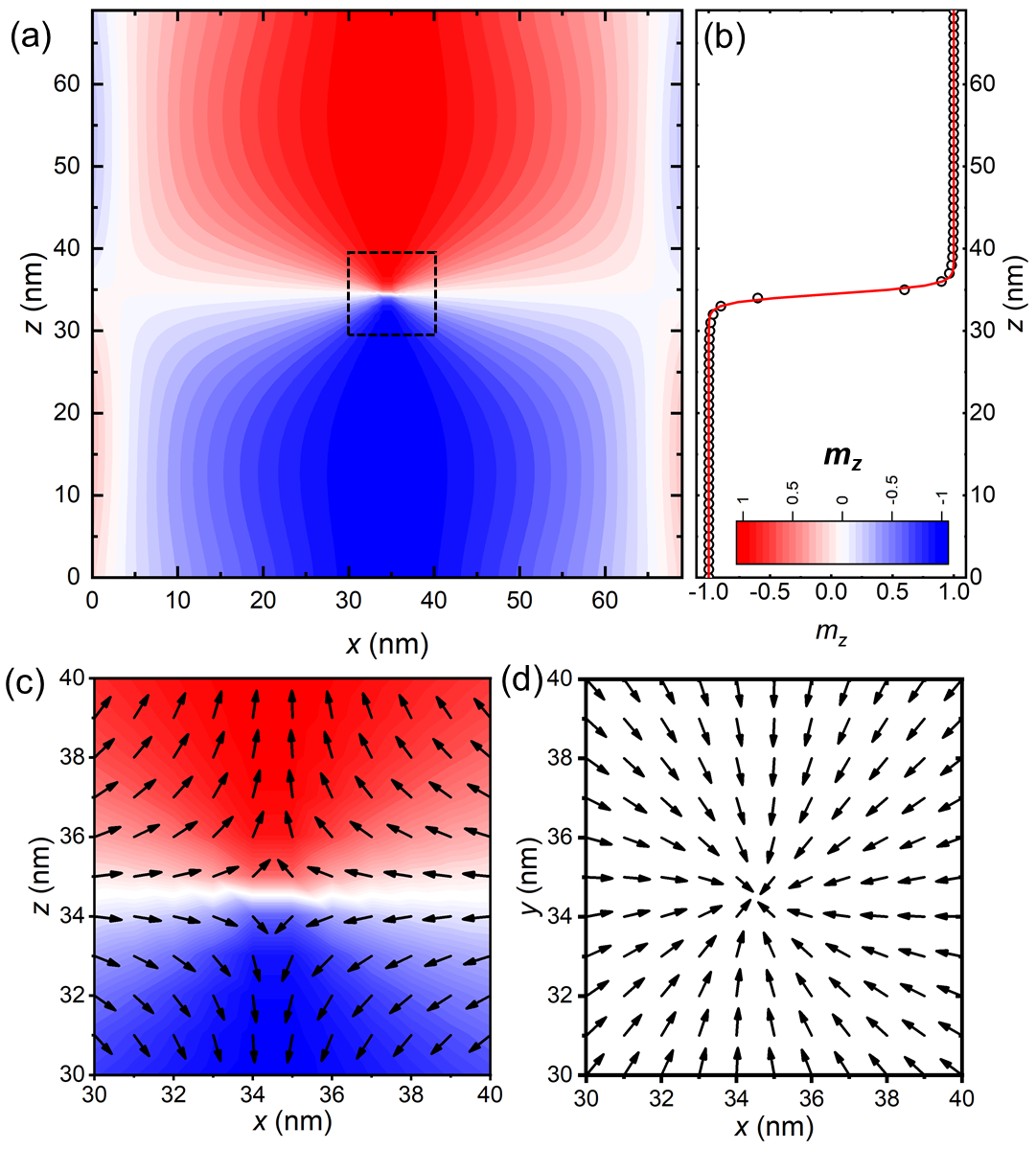} 
\caption{Contour plot of the magnetization $z$-component in the $xz$ cubic plane showing the Bloch point in the center of the cube. The magnetization profile can be described by a Bloch wall with $m_z=\tanh(z/\delta_\mathrm{dw})$ with a width $\delta_\mathrm{dw}$ of exactly one simulation cell (here 1 nm). Detailed contour and vector plot of (b) the $xz$ and (c) the $xy$ plane in the vicinity of the Bloch point, showing how the center resides at simulation-cell boundaries. }
\label{Profile}
\end{figure}

The resulting configuration, as shown in Fig. \ref{MM_Main} for a 70 nm $\times$ 70 nm $\times$ 70 nm cube, corresponds to a single $\pi$ rotation of the local magnetization in all directions. This state consists of two skyrmionic textures on the top and bottom surfaces of the cuboid, extending below the surfaces in the form of chiral bobbers. Because the winding of the top ChB is exactly opposite to that of the bottom ChB, the twisting of the magnetization around the point where the ChBs meet is a topological point-defect, i.e. the Bloch-point singularity. The BP is created by and confined between the two ChBs. 

Figure \ref{Profile} shows a contour plot of the $z$ component of the magnetization in the $xz$ plane and the magnetization profile along the $z$ axis. The profile corresponds to that of a $\pi$ domain wall with a width of exactly one simulation cell (here one nanometer). The same profile is found for $m_x$ and $m_y$, as illustrated in Fig. \ref{Profile}(d). At the center of the BP the magnetization vanishes, therefore the point of zero $\mathbf{m}$ cannot reside in any single simulation cell: it is always located between simulation cells. 

Considering the extremely steep magnetization gradient close to the BP, the singularity itself is beyond the micromagnetic continuum approximation that assumes a smoothly varying magnetization, and atomistic modeling has been pursued \cite{andreas2014} for a detailed analysis of the region close to the BP center. As it will be pointed out below, however, the BP carries a substantial magnetic volume around it and capturing the behavior of the magnetization around the singularity \cite{hubert1998} is what provides the insight to the BP itself. Additionally, a recent comparison between micromagnetics and atomistic simulations indicated that the two approximations come to a quantitative agreement when the micromagnetic simulation cell is comparable to the lattice constant \cite{yuli2019}.

Importantly, whether or not a BP is created, depends on the initial state of the magnetization, i.e., whether the system is initialized in a fully demagnetized state or in a fully polarized state. The key aspect here is therefore the path followed by the magnetization winding process. If the initial state is fully random the equilibrium state will be that of a skyrmion tube or a helicoid, both of which have a slightly lower energy than the BP state. However, if the system is initialized with a uniform ferromagnetic state, corresponding to the application of a saturating field in an experiment, the Bloch point is created through the process described above and is stabilized by the topological constraints on the magnetization texture. Similar history-dependence was also predicted for cylindrical FeGe nanowires \cite{charilaou2017}.

To quantify the magnetization winding around the BP, the winding number, or topological charge $Q$, is evaluated on the surface of a sphere $S^2$ around the BP with \cite{malozemoff1979}
\begin{equation}
Q_{S^2} =\frac{1}{4\pi} \int_{S^2} \! d\vartheta d\varphi \,(\partial_\vartheta \theta \partial_\varphi \phi -  \partial_\varphi \theta \partial_\vartheta \phi) \sin \theta \; ,
\end{equation}
where $\vartheta,\varphi$ parametrize the surface around the BP at a radius $r$ with $\mathbf{r}=( r\sin \vartheta \cos \varphi, r\sin \vartheta \sin \varphi, r\cos \vartheta)$. In the vector profiles of the $xz$ and the $xy$ planes around the BP, shown in Fig. \ref{Profile}(c,d), the magnetization follows $\theta(\vartheta)=-\vartheta$ and $\phi(\varphi)=\pi+\varphi$, yielding $Q_{S^2}  =+1$, i.e., the confined BP found here is topologically equivalent to the radially outward-pointing hedgehog BP \cite{pylypovski2012} (see Fig. \ref{BPoints}a), and it has the same configuration as those discussed in Ref. \cite{kanazawa2017} for MnGe, in \cite{tokura2019} for MnSi$_{1-x}$Ge$_{x}$, and in \cite{im2019} for Fe$_{20}$Ni$_{80}$. It should be noted that for opposite DMi sign, the inverse configuration occurs (see Fig. \ref{BPoints}d).
\begin{figure}
	\centering
\includegraphics[width=1.0\columnwidth]{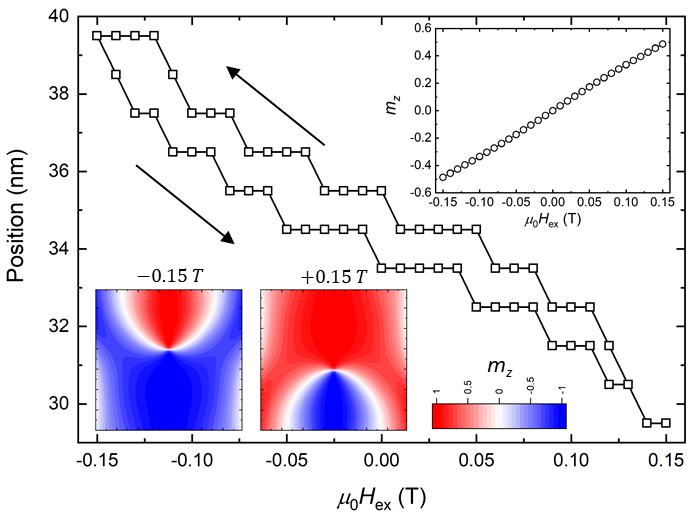} 
\caption{Position of the Bloch point as a function of external field showing discreet jumps between neighboring cell-boundaries. The upper-right inset shows the magnetization as a function of external field, which follows a linear trend. The lower-left insets show contour plots of $m_z$ in the $xz$ plane at (left) $\mu_0 H = -0.15$ T and (right) $\mu_0 H = +0.15$ T, where the Bloch point center is pushed by the growing region with magnetization parallel to the applied field.   }
\label{Motion}
\end{figure}

Further, application of an external field moves the BP inside the material without destroying it. As shown in Fig. \ref{Motion}, the position of the BP changes with increasing field strength with discreet jumps to neighboring cell-boundaries. The motion has a signature of field-dragging, which is associated with mesh-friction \cite{thiaville2003} and depends on the size of the simulation cell, as discussed in Ref. \cite{yuli2019}. The motion only takes place if the external field strength exceeds a value of 45 mT.

Depending on the direction of the external field, the region magnetized parallel to the field grows with increasing field strength, similar to classical domain behavior \cite{kittel1949}, which consequently pushes the BP in the opposite direction. With increasing field strength the BP moves further away from its equilibrium position, and with decreasing field strength it returns back to its original position. The net magnetization of the cuboid changes linearly with the external field and exhibits zero hysteresis, therefore this process is fully reversible and the BP can be moved up and down reversibly and repeatedly. 

It is notable how the BP remains stable against an external magnetic field of moderate strength. Only when the external field exceeds a critical value (here $\mu_0 H_\mathrm{ex}>160$ mT) does the BP become expelled from the system through one of the surfaces and the system transforms to the uniform FM state. Once the system is in the FM state, however, upon removing the field a new BP will be formed with the same process described above. 

\begin{figure}
	\centering
\includegraphics[width=0.99\columnwidth]{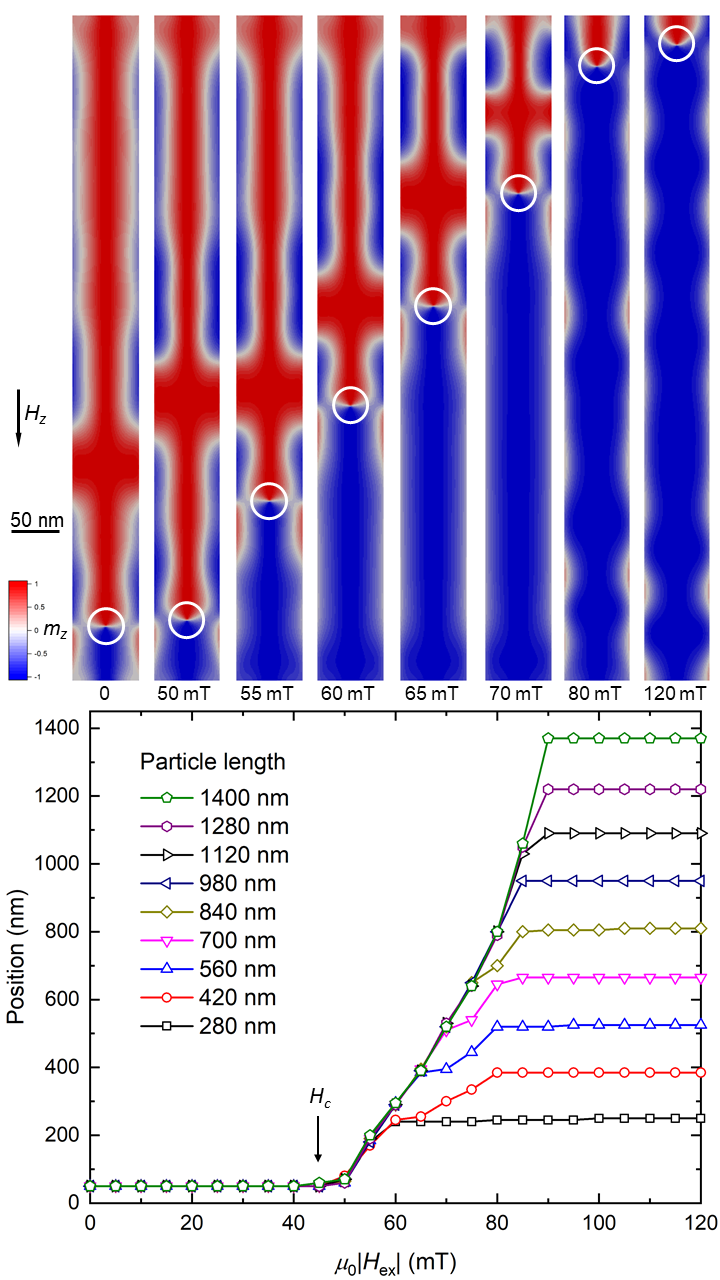} 
\caption{Contour plots of the $z$ component of the magnetization vector showing the field-driven displacement of the Bloch Point with increasing external field-strength, and (bottom panel) comparison between the BP motion in structures with different aspect ratios. For all structures the motion of the BP starts at an external field strength of 45 mT and the displacement follows a nearly linear trend until it reaches the other end of the structure.}
\label{BP_Drive}
\end{figure}
There results were obtained for a cubic structure that can fit exactly one complete period of the magnetization texture, i.e., all sides have $L=\lambda$. Structures with higher aspect ratio exhibit very similar behavior, with the main difference that they provide more vertical space for the motion of the BP. To test this, structures with an aspect ratio of up to 20 (70 nm $\times$ 70 nm $\times$ 1400 nm) were simulated. One example for a structure with aspect ratio 10 is illustrated in Fig. \ref{BP_Drive} which shows simulation snapshots as contour plots of $m_z$ in the $yz$ plane with increasing external field-strength. The bottom panel of Fig. \ref{BP_Drive} compares the motion of the Bloch Point with increasing external field-strength for structures with different length. For all structures the BP is pinned to its equilibrium position for external field-strengths of up to 45 mT, which can be defined as the depinning field. Upon further increase of the field strength, the BP starts propagating upwards, in the direction opposing the field, until it approaches the other end of the structure. Further increase of $H_z$ slightly pushes the BP closer to the end of the structure with a minimum distance between BP center and surface of 30 nm. Finally, at an external field of 160 mT the BP becomes expelled through the surface, as discussed above.

Surprisingly, the depinning field-strength is exactly the same for all structures. The pinning can be attributed to intrinsic contributions, due to the local magnetic configuration and the associated exchange and DMi energy, and to extrinsic contributions, i.e., the magnetostatics of free surfaces. The maximum pinning force as discussed in Ref. \cite{kim2013}, is $F_\mathrm{p}=2\pi c A$, where $c\approx 1$ is a numerical constant, and the depinning field $H_c$ depends on the saturation magnetization and the physical size of the BP, a sphere with radius $R$: $\mu_0 H_c = c A/ M_\mathrm{s} R^2 = c\mu_0 M_\mathrm{s}\delta_\mathrm{ex}^2/2R^2$. Setting this field strength equal to 45 mT, as found in the simulations, yields an effective BP radius of $R\approx 23$ nm. This radius is in good agreement with the minimum distance between BP center and the surface (30 nm), and it reveals an important aspect of the BP texture: even though the BP is by definition zero-dimensional, the resulting texture around it encloses a substantial magnetic volume. This also highlights how micromagnetic simulations are suited to capture the behavior of the BP by modeling the texture around it.

To further support the micromagnetic findings, atomistic simulations were performed with the software package SPIRIT \cite{spirit}. The Heisenberg-type Hamiltonian contains the atomistic analogs of the same contributions as in equation \ref{micromag_en}, consisting of the nearest-neighbor ferromagnetic exchange interaction energy $J$, the DMi $\mathbf{D}$, the external field, and dipole-dipole interactions.
\begin{align}
\mathcal{H} =& - J\sum_{ij} \textbf{S}_i \cdot \textbf{S}_j - \sum_{ij} \textbf{D} \cdot \left(\textbf{S}_i \times \textbf{S}_j\right)-\mu_0\sum_{i}\mathbf{H}_\mathrm{ex}\cdot \textbf{S}_i \nonumber \\
&+\frac{1}{2}\frac{\mu_0 S^2}{4\pi}\sum_{ij} \frac{\left(\textbf{S}_i\cdot\hat{\textbf{r}}_{ij} \right)\left(\textbf{S}_j\cdot \hat{\textbf{r}}_{ij} \right) -\textbf{S}_i\cdot \textbf{S}_j}{r_{ij}^3} \; .
\end{align} where $S=1$ is the atomic magnetic moment and $\textbf{r}_{ij}$ is the vector connecting $\textbf{S}_i$ and $\textbf{S}_j$.

For computational efficiency, the FeGe structure was mapped onto a simple cubic lattice. The nearest-neighbor exchange interaction was derived from the Curie temperature ($T_\mathrm{C}=280$ K) as $J=T_\mathrm{C}k_\mathrm{B}/n\left<S\right>^2\approx 24$ meV, where $n=1$ is the number of atoms in the unit cell. The DMi strength was adjusted so that $\lambda$ is equal to the size of the system. A cube of FeGe with a side length of 70 nm contains about $10^8$ atoms, which is computationally challenging, therefore smaller systems were simulated and the interaction parameters were scaled ($J\rightarrow \hat{J}$, $D\rightarrow \hat{D}$ ) so that the dipole-dipole interaction strength is equivalent to that for a 70 nm cube. This was done by adjusting the exchange length $\delta\propto \sqrt{J} \rightarrow \hat\delta\propto\sqrt{\hat{J}}$, where for appropriate scaling $\hat{\delta}$ needs to be reduced by the same fraction as the size of the system: $\hat{\delta}/\delta=\hat{L}/L$, and therefore $\hat{J}=J(\hat{L}/L)^2$. For example, a cubic lattice with a side of 24 atoms has $\hat{J}=J(24a/70)^2\approx 0.64$ meV and the corresponding DMi strength, for which $\lambda$ is equal to the system size, is $\hat{D}=0.17$ meV. The parameter scaling was tested with several lattice sizes ($L=16$, $20$, $24$, and $32$) and it was found to reproduce the same result each time. 

\begin{figure}
	\centering
\includegraphics[width=0.95\columnwidth]{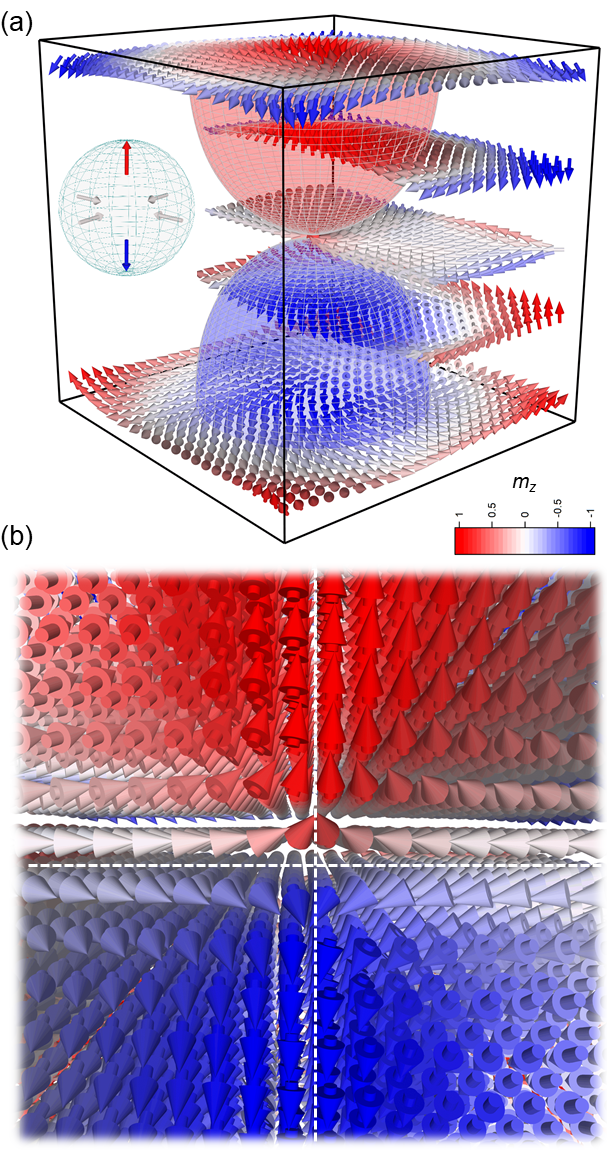} 
\caption{Atomistic spin dynamics simulations of a system with $L=24$ showing the same configuration as that predicted by micromagnetic simulations in Fig. \ref{MM_Main} (a) where a Bloch point is confined in between two chiral bobbers. Panel (b) shows a detailed view of the magnetization configuration close to the BP. The intersection of the dashed lines indicates the center of the BP.}
\label{ASD_Main}
\end{figure}
Figure \ref{ASD_Main} shows the results of the atomistic simulations for a system with $L=24$, which are in excellent agreement with the findings from the micromagnetic simulations: starting from a uniform FM configuration, the surface moments curl in a process that leads to an individual BP confined at the intersection between two ChBs with opposite polarity. Moreover, in the same way that the magnetization of a micromagnetic cell cannot vanish, the magnetic moment of an atom cannot be zero, therefore the center of the BP is always between two neighboring atoms. Application of an external field drives the BP up and down the structure, where the motion is again in discreet jumps to neighboring inter-atomic locations (not shown here). The atomistic simulations therefore confirm the micromagnetic findings at zero temperature, regarding both the formation process and the field-driven motion of the confined BP. 

Turning to the energetics, the energy of the BP state described above is very close to that of a helicoid texture and a skyrmion-tube texture. In the examples shown here the total energy of the BP state is actually $\sim 1\%$ higher than that of the skyrmionic state, but the energy penalty of a few $\upmu$eV/atom is not enough for the system to overcome the topological constraints and transform to the skyrmion configuration or the helicoid configuration at zero temperature. Atomistic simulations at finite-temperature, where the temperature was implemented as a random thermal field \cite{spirit}, show that the BP becomes unstable and escapes through the cuboid's edges, but only for temperatures above $\sim$30\% of the Curie temperature $T_\mathrm{C}$. Hence, for FeGe with $T_\mathrm{C}=280$ K, these findings suggest that it should be experimentally possible to observe the creation of the confined BP at liquid nitrogen temperatures.

A proposed experimental protocol for the observation of the predicted phenomena is to perform field-cooling of a FeGe cuboid of comparable dimensionality below $T_\mathrm{C}/3$ to initialize the system in a FM configuration and then turn off the external field. Then, variation of the external field within the order of 100 mT will lead to the motion of the BP inside the structure. It should be noted, however, that the simulations assume a perfect and defect-free structure, and therefore the depinning field and the annihilation field are expected to be reduced in an experiment due to the presence of roughness and/or defects. 

\section{Conclusions}
In summary, this study predicts that individual Bloch points can be stabilized in magnetic nanocuboids by magnetostatic and Dzyaloshinskii-Moriya interactions. The formation of the Bloch point is initiated by the simultaneous winding of two skyrmionic textures with opposite polarity on the top and bottom surface of the cube that form a pair of opposing chiral bobbers. The simulations predict that the Bloch point remains stable against an external magnetic field with moderate strength and that the external field can move the Bloch point inside the material when the field strength exceeds a depinning field of 45 mT. The stability of the Bloch point is remarkable and its confinement between two chiral bobbers illustrates the diversity of topological objects obtainable in nanostructures, enabled in tandem by intrinsic and extrinsic magnetic forces. The precise control of topological point-defects described here can open the door to a new range of experiments to study the elusive magnetic singularity and to evaluate the emergent electrodynamics of moving topological textures.

\begin{acknowledgments}
The author is grateful to Leonardo Pierobon, Hans-Benjamin Braun and Rafal Dunin-Borkowski for fruitful discussions.
\end{acknowledgments}
%\bibliographystyle{apsrev4-1}
%\bibliography{Literature}{ }

\begin{thebibliography}{52}%
\makeatletter
\providecommand \@ifxundefined [1]{%
 \@ifx{#1\undefined}
}%
\providecommand \@ifnum [1]{%
 \ifnum #1\expandafter \@firstoftwo
 \else \expandafter \@secondoftwo
 \fi
}%
\providecommand \@ifx [1]{%
 \ifx #1\expandafter \@firstoftwo
 \else \expandafter \@secondoftwo
 \fi
}%
\providecommand \natexlab [1]{#1}%
\providecommand \enquote  [1]{``#1''}%
\providecommand \bibnamefont  [1]{#1}%
\providecommand \bibfnamefont [1]{#1}%
\providecommand \citenamefont [1]{#1}%
\providecommand \href@noop [0]{\@secondoftwo}%
\providecommand \href [0]{\begingroup \@sanitize@url \@href}%
\providecommand \@href[1]{\@@startlink{#1}\@@href}%
\providecommand \@@href[1]{\endgroup#1\@@endlink}%
\providecommand \@sanitize@url [0]{\catcode `\\12\catcode `\$12\catcode
  `\&12\catcode `\#12\catcode `\^12\catcode `\_12\catcode `\%12\relax}%
\providecommand \@@startlink[1]{}%
\providecommand \@@endlink[0]{}%
\providecommand \url  [0]{\begingroup\@sanitize@url \@url }%
\providecommand \@url [1]{\endgroup\@href {#1}{\urlprefix }}%
\providecommand \urlprefix  [0]{URL }%
\providecommand \Eprint [0]{\href }%
\providecommand \doibase [0]{http://dx.doi.org/}%
\providecommand \selectlanguage [0]{\@gobble}%
\providecommand \bibinfo  [0]{\@secondoftwo}%
\providecommand \bibfield  [0]{\@secondoftwo}%
\providecommand \translation [1]{[#1]}%
\providecommand \BibitemOpen [0]{}%
\providecommand \bibitemStop [0]{}%
\providecommand \bibitemNoStop [0]{.\EOS\space}%
\providecommand \EOS [0]{\spacefactor3000\relax}%
\providecommand \BibitemShut  [1]{\csname bibitem#1\endcsname}%
\let\auto@bib@innerbib\@empty
%</preamble>
\bibitem [{\citenamefont {Wang}\ and\ \citenamefont {Zhang}(2017)}]{wang2017}%
  \BibitemOpen
  \bibfield  {author} {\bibinfo {author} {\bibfnamefont {J.}~\bibnamefont
  {Wang}}\ and\ \bibinfo {author} {\bibfnamefont {S.-C.}\ \bibnamefont
  {Zhang}},\ }\href@noop {} {\bibfield  {journal} {\bibinfo  {journal} {Nature
  Mat.}\ }\textbf {\bibinfo {volume} {16}},\ \bibinfo {pages} {1062} (\bibinfo
  {year} {2017})}\BibitemShut {NoStop}%
\bibitem [{\citenamefont {Braun}(2012)}]{braun2012}%
  \BibitemOpen
  \bibfield  {author} {\bibinfo {author} {\bibfnamefont {H.-B.}\ \bibnamefont
  {Braun}},\ }\href@noop {} {\bibfield  {journal} {\bibinfo  {journal} {Adv.
  Phys.}\ }\textbf {\bibinfo {volume} {61}},\ \bibinfo {pages} {1} (\bibinfo
  {year} {2012})}\BibitemShut {NoStop}%
\bibitem [{\citenamefont {Ackerman}\ and\ \citenamefont
  {Smalyukh}(2017)}]{ackerman2017}%
  \BibitemOpen
  \bibfield  {author} {\bibinfo {author} {\bibfnamefont {P.~J.}\ \bibnamefont
  {Ackerman}}\ and\ \bibinfo {author} {\bibfnamefont {I.~I.}\ \bibnamefont
  {Smalyukh}},\ }\href@noop {} {\bibfield  {journal} {\bibinfo  {journal}
  {Nature Mat.}\ }\textbf {\bibinfo {volume} {16}},\ \bibinfo {pages} {426}
  (\bibinfo {year} {2017})}\BibitemShut {NoStop}%
\bibitem [{\citenamefont {Zang}\ \emph {et~al.}(2018)\citenamefont {Zang},
  \citenamefont {Cros},\ and\ \citenamefont {Hoffmann}}]{zang2018}%
  \BibitemOpen
  \bibinfo {editor} {\bibfnamefont {J.}~\bibnamefont {Zang}}, \bibinfo {editor}
  {\bibfnamefont {V.}~\bibnamefont {Cros}}, \ and\ \bibinfo {editor}
  {\bibfnamefont {A.}~\bibnamefont {Hoffmann}},\ eds.,\ \href@noop {} {\emph
  {\bibinfo {title} {Topology in Magnetism}}}\ (\bibinfo  {publisher}
  {Springer},\ \bibinfo {year} {2018})\BibitemShut {NoStop}%
\bibitem [{\citenamefont {R\"ossler}\ \emph {et~al.}(2006)\citenamefont
  {R\"ossler}, \citenamefont {Bogdanov},\ and\ \citenamefont
  {Pfleiderer}}]{roessler2006}%
  \BibitemOpen
  \bibfield  {author} {\bibinfo {author} {\bibfnamefont {U.~K.}\ \bibnamefont
  {R\"ossler}}, \bibinfo {author} {\bibfnamefont {A.~N.}\ \bibnamefont
  {Bogdanov}}, \ and\ \bibinfo {author} {\bibfnamefont {C.}~\bibnamefont
  {Pfleiderer}},\ }\href@noop {} {\bibfield  {journal} {\bibinfo  {journal}
  {Nature}\ }\textbf {\bibinfo {volume} {797--801}},\ \bibinfo {pages} {442}
  (\bibinfo {year} {2006})}\BibitemShut {NoStop}%
\bibitem [{\citenamefont {Hellman}\ \emph {et~al.}(2017)\citenamefont
  {Hellman}, \citenamefont {Hoffmann}, \citenamefont {Tserkovnyak},
  \citenamefont {Beach}, \citenamefont {Fullerton}, \citenamefont {Leighton},
  \citenamefont {MacDonald}, \citenamefont {Ralph}, \citenamefont {Arena},
  \citenamefont {DÃ¼rr}, \citenamefont {Fischer}, \citenamefont {Grollier},
  \citenamefont {Heremans}, \citenamefont {Jungwirth}, \citenamefont {Kimel},
  \citenamefont {Koopmans}, \citenamefont {Krivorotov}, \citenamefont {May},
  \citenamefont {Petford-Long}, \citenamefont {Rondinelli}, \citenamefont
  {Samarth}, \citenamefont {Schuller}, \citenamefont {Slavin}, \citenamefont
  {Stiles}, \citenamefont {Tchernyshyov}, \citenamefont {Thiaville},\ and\
  \citenamefont {Zink}}]{hellman2017}%
  \BibitemOpen
  \bibfield  {author} {\bibinfo {author} {\bibfnamefont {F.}~\bibnamefont
  {Hellman}}, \bibinfo {author} {\bibfnamefont {A.}~\bibnamefont {Hoffmann}},
  \bibinfo {author} {\bibfnamefont {Y.}~\bibnamefont {Tserkovnyak}}, \bibinfo
  {author} {\bibfnamefont {G.~S.}\ \bibnamefont {Beach}}, \bibinfo {author}
  {\bibfnamefont {E.~E.}\ \bibnamefont {Fullerton}}, \bibinfo {author}
  {\bibfnamefont {C.}~\bibnamefont {Leighton}}, \bibinfo {author}
  {\bibfnamefont {A.~H.}\ \bibnamefont {MacDonald}}, \bibinfo {author}
  {\bibfnamefont {D.~C.}\ \bibnamefont {Ralph}}, \bibinfo {author}
  {\bibfnamefont {D.~A.}\ \bibnamefont {Arena}}, \bibinfo {author}
  {\bibfnamefont {H.~A.}\ \bibnamefont {D\"urr}}, \bibinfo {author}
  {\bibfnamefont {P.}~\bibnamefont {Fischer}}, \bibinfo {author} {\bibfnamefont
  {J.}~\bibnamefont {Grollier}}, \bibinfo {author} {\bibfnamefont {J.~P.}\
  \bibnamefont {Heremans}}, \bibinfo {author} {\bibfnamefont {T.}~\bibnamefont
  {Jungwirth}}, \bibinfo {author} {\bibfnamefont {A.~V.}\ \bibnamefont
  {Kimel}}, \bibinfo {author} {\bibfnamefont {B.}~\bibnamefont {Koopmans}},
  \bibinfo {author} {\bibfnamefont {I.~N.}\ \bibnamefont {Krivorotov}},
  \bibinfo {author} {\bibfnamefont {S.~J.}\ \bibnamefont {May}}, \bibinfo
  {author} {\bibfnamefont {A.~K.}\ \bibnamefont {Petford-Long}}, \bibinfo
  {author} {\bibfnamefont {J.~M.}\ \bibnamefont {Rondinelli}}, \bibinfo
  {author} {\bibfnamefont {N.}~\bibnamefont {Samarth}}, \bibinfo {author}
  {\bibfnamefont {I.~K.}\ \bibnamefont {Schuller}}, \bibinfo {author}
  {\bibfnamefont {A.~N.}\ \bibnamefont {Slavin}}, \bibinfo {author}
  {\bibfnamefont {M.~D.}\ \bibnamefont {Stiles}}, \bibinfo {author}
  {\bibfnamefont {O.}~\bibnamefont {Tchernyshyov}}, \bibinfo {author}
  {\bibfnamefont {A.}~\bibnamefont {Thiaville}}, \ and\ \bibinfo {author}
  {\bibfnamefont {B.~L.}\ \bibnamefont {Zink}},\ }\href {\doibase
  10.1103/revmodphys.89.025006} {\bibfield  {journal} {\bibinfo  {journal}
  {Rev. Mod. Phys.}\ }\textbf {\bibinfo {volume} {89}},\ \bibinfo {pages}
  {025006} (\bibinfo {year} {2017})}\BibitemShut {NoStop}%
\bibitem [{\citenamefont {Fert}\ \emph {et~al.}(2017)\citenamefont {Fert},
  \citenamefont {Reyren},\ and\ \citenamefont {Cros}}]{fert2017}%
  \BibitemOpen
  \bibfield  {author} {\bibinfo {author} {\bibfnamefont {A.}~\bibnamefont
  {Fert}}, \bibinfo {author} {\bibfnamefont {N.}~\bibnamefont {Reyren}}, \ and\
  \bibinfo {author} {\bibfnamefont {V.}~\bibnamefont {Cros}},\ }\href@noop {}
  {\bibfield  {journal} {\bibinfo  {journal} {Nature Rev. Mat.}\ }\textbf
  {\bibinfo {volume} {2}},\ \bibinfo {pages} {17031} (\bibinfo {year}
  {2017})}\BibitemShut {NoStop}%
\bibitem [{\citenamefont {Jiang}\ \emph {et~al.}(2017)\citenamefont {Jiang},
  \citenamefont {Chen}, \citenamefont {Liu}, \citenamefont {Zang},
  \citenamefont {te~Velthuis},\ and\ \citenamefont {Hoffmann}}]{hoffmann2017}%
  \BibitemOpen
  \bibfield  {author} {\bibinfo {author} {\bibfnamefont {W.}~\bibnamefont
  {Jiang}}, \bibinfo {author} {\bibfnamefont {G.}~\bibnamefont {Chen}},
  \bibinfo {author} {\bibfnamefont {K.}~\bibnamefont {Liu}}, \bibinfo {author}
  {\bibfnamefont {J.}~\bibnamefont {Zang}}, \bibinfo {author} {\bibfnamefont
  {S.~G.}\ \bibnamefont {te~Velthuis}}, \ and\ \bibinfo {author} {\bibfnamefont
  {A.}~\bibnamefont {Hoffmann}},\ }\href@noop {} {\bibfield  {journal}
  {\bibinfo  {journal} {Phys. Rep.}\ }\textbf {\bibinfo {volume} {704}},\
  \bibinfo {pages} {1} (\bibinfo {year} {2017})}\BibitemShut {NoStop}%
\bibitem [{\citenamefont {Everschor-Sitte}\ \emph {et~al.}(2018)\citenamefont
  {Everschor-Sitte}, \citenamefont {Masell}, \citenamefont {Reeve},\ and\
  \citenamefont {Kl\"aui}}]{everschor2018}%
  \BibitemOpen
  \bibfield  {author} {\bibinfo {author} {\bibfnamefont {K.}~\bibnamefont
  {Everschor-Sitte}}, \bibinfo {author} {\bibfnamefont {J.}~\bibnamefont
  {Masell}}, \bibinfo {author} {\bibfnamefont {R.~M.}\ \bibnamefont {Reeve}}, \
  and\ \bibinfo {author} {\bibfnamefont {M.}~\bibnamefont {Kl\"aui}},\
  }\href@noop {} {\bibfield  {journal} {\bibinfo  {journal} {J. App. Phys.}\
  }\textbf {\bibinfo {volume} {124}},\ \bibinfo {pages} {240901} (\bibinfo
  {year} {2018})}\BibitemShut {NoStop}%
\bibitem [{\citenamefont {Koshibae}\ and\ \citenamefont
  {Nagaosa}(2016)}]{koshibae2016}%
  \BibitemOpen
  \bibfield  {author} {\bibinfo {author} {\bibfnamefont {W.}~\bibnamefont
  {Koshibae}}\ and\ \bibinfo {author} {\bibfnamefont {N.}~\bibnamefont
  {Nagaosa}},\ }\href@noop {} {\bibfield  {journal} {\bibinfo  {journal}
  {Nature Comm.}\ }\textbf {\bibinfo {volume} {7}},\ \bibinfo {pages} {10542}
  (\bibinfo {year} {2016})}\BibitemShut {NoStop}%
\bibitem [{\citenamefont {Rybakov}\ \emph {et~al.}(2015)\citenamefont
  {Rybakov}, \citenamefont {Borisov}, \citenamefont {Bl\"ugel},\ and\
  \citenamefont {Kiselev}}]{kiselev2015}%
  \BibitemOpen
  \bibfield  {author} {\bibinfo {author} {\bibfnamefont {F.~N.}\ \bibnamefont
  {Rybakov}}, \bibinfo {author} {\bibfnamefont {A.~B.}\ \bibnamefont
  {Borisov}}, \bibinfo {author} {\bibfnamefont {S.}~\bibnamefont {Bl\"ugel}}, \
  and\ \bibinfo {author} {\bibfnamefont {N.~S.}\ \bibnamefont {Kiselev}},\
  }\href@noop {} {\bibfield  {journal} {\bibinfo  {journal} {Phys. Rev. Lett.}\
  }\textbf {\bibinfo {volume} {115}},\ \bibinfo {pages} {117201} (\bibinfo
  {year} {2015})}\BibitemShut {NoStop}%
\bibitem [{\citenamefont {Zheng}\ \emph {et~al.}(2018)\citenamefont {Zheng},
  \citenamefont {Rybakov}, \citenamefont {Borisov}, \citenamefont {Song},
  \citenamefont {Wang}, \citenamefont {Li}, \citenamefont {Du}, \citenamefont
  {Kiselev}, \citenamefont {Caron}, \citenamefont {Kovacs}, \citenamefont
  {Tian}, \citenamefont {Zhang}, \citenamefont {Bl\"ugel},\ and\ \citenamefont
  {Dunin-Borkowski}}]{kiselev2018}%
  \BibitemOpen
  \bibfield  {author} {\bibinfo {author} {\bibfnamefont {F.}~\bibnamefont
  {Zheng}}, \bibinfo {author} {\bibfnamefont {F.~N.}\ \bibnamefont {Rybakov}},
  \bibinfo {author} {\bibfnamefont {A.~B.}\ \bibnamefont {Borisov}}, \bibinfo
  {author} {\bibfnamefont {D.}~\bibnamefont {Song}}, \bibinfo {author}
  {\bibfnamefont {S.}~\bibnamefont {Wang}}, \bibinfo {author} {\bibfnamefont
  {Z.-A.}\ \bibnamefont {Li}}, \bibinfo {author} {\bibfnamefont
  {H.}~\bibnamefont {Du}}, \bibinfo {author} {\bibfnamefont {N.~S.}\
  \bibnamefont {Kiselev}}, \bibinfo {author} {\bibfnamefont {J.}~\bibnamefont
  {Caron}}, \bibinfo {author} {\bibfnamefont {A.}~\bibnamefont {Kovacs}},
  \bibinfo {author} {\bibfnamefont {M.}~\bibnamefont {Tian}}, \bibinfo {author}
  {\bibfnamefont {Y.}~\bibnamefont {Zhang}}, \bibinfo {author} {\bibfnamefont
  {S.}~\bibnamefont {Bl\"ugel}}, \ and\ \bibinfo {author} {\bibfnamefont
  {R.~E.}\ \bibnamefont {Dunin-Borkowski}},\ }\href@noop {} {\bibfield
  {journal} {\bibinfo  {journal} {Nature Nano.}\ }\textbf {\bibinfo {volume}
  {13}},\ \bibinfo {pages} {451} (\bibinfo {year} {2018})}\BibitemShut
  {NoStop}%
\bibitem [{\citenamefont {Fern\'andez-Pacheco}\ \emph
  {et~al.}(2017)\citenamefont {Fern\'andez-Pacheco}, \citenamefont {Streubel},
  \citenamefont {Fruchart}, \citenamefont {Hertel}, \citenamefont {Fischer},\
  and\ \citenamefont {Cowburn}}]{pacheco2017}%
  \BibitemOpen
  \bibfield  {author} {\bibinfo {author} {\bibfnamefont {A.}~\bibnamefont
  {Fern\'andez-Pacheco}}, \bibinfo {author} {\bibfnamefont {R.}~\bibnamefont
  {Streubel}}, \bibinfo {author} {\bibfnamefont {O.}~\bibnamefont {Fruchart}},
  \bibinfo {author} {\bibfnamefont {R.}~\bibnamefont {Hertel}}, \bibinfo
  {author} {\bibfnamefont {P.}~\bibnamefont {Fischer}}, \ and\ \bibinfo
  {author} {\bibfnamefont {R.~P.}\ \bibnamefont {Cowburn}},\ }\href@noop {}
  {\bibfield  {journal} {\bibinfo  {journal} {Nature Comm.}\ }\textbf {\bibinfo
  {volume} {8}},\ \bibinfo {pages} {15756} (\bibinfo {year}
  {2017})}\BibitemShut {NoStop}%
\bibitem [{\citenamefont {Sampaio}\ \emph {et~al.}(2013)\citenamefont
  {Sampaio}, \citenamefont {Cros}, \citenamefont {Rohart}, \citenamefont
  {Thiaville},\ and\ \citenamefont {Fert}}]{sampaio2013}%
  \BibitemOpen
  \bibfield  {author} {\bibinfo {author} {\bibfnamefont {J.}~\bibnamefont
  {Sampaio}}, \bibinfo {author} {\bibfnamefont {V.}~\bibnamefont {Cros}},
  \bibinfo {author} {\bibfnamefont {S.}~\bibnamefont {Rohart}}, \bibinfo
  {author} {\bibfnamefont {A.}~\bibnamefont {Thiaville}}, \ and\ \bibinfo
  {author} {\bibfnamefont {A.}~\bibnamefont {Fert}},\ }\href@noop {} {\bibfield
   {journal} {\bibinfo  {journal} {Nature Nano.}\ }\textbf {\bibinfo {volume}
  {8}},\ \bibinfo {pages} {839} (\bibinfo {year} {2013})}\BibitemShut {NoStop}%
\bibitem [{\citenamefont {Fert}\ \emph {et~al.}(2013)\citenamefont {Fert},
  \citenamefont {Cross},\ and\ \citenamefont {Sampaio}}]{fert2013}%
  \BibitemOpen
  \bibfield  {author} {\bibinfo {author} {\bibfnamefont {A.}~\bibnamefont
  {Fert}}, \bibinfo {author} {\bibfnamefont {V.}~\bibnamefont {Cross}}, \ and\
  \bibinfo {author} {\bibfnamefont {J.}~\bibnamefont {Sampaio}},\ }\href@noop
  {} {\bibfield  {journal} {\bibinfo  {journal} {Nature Nano.}\ }\textbf
  {\bibinfo {volume} {8}},\ \bibinfo {pages} {152} (\bibinfo {year}
  {2013})}\BibitemShut {NoStop}%
\bibitem [{\citenamefont {Moreau-Luchaire}\ \emph {et~al.}(2016)\citenamefont
  {Moreau-Luchaire}, \citenamefont {Moutafis}, \citenamefont {Reyren},
  \citenamefont {Sampaio}, \citenamefont {Vaz}, \citenamefont {Horne},
  \citenamefont {Bouzehouane}, \citenamefont {Garcia}, \citenamefont
  {Deranlot}, \citenamefont {Warnicke}, \citenamefont {Wohlh\"uter},
  \citenamefont {George}, \citenamefont {Weigand}, \citenamefont {Raabe},
  \citenamefont {Cros},\ and\ \citenamefont {Fert}}]{moutafis2016}%
  \BibitemOpen
  \bibfield  {author} {\bibinfo {author} {\bibfnamefont {C.}~\bibnamefont
  {Moreau-Luchaire}}, \bibinfo {author} {\bibfnamefont {C.}~\bibnamefont
  {Moutafis}}, \bibinfo {author} {\bibfnamefont {N.}~\bibnamefont {Reyren}},
  \bibinfo {author} {\bibfnamefont {J.}~\bibnamefont {Sampaio}}, \bibinfo
  {author} {\bibfnamefont {C.~A.~F.}\ \bibnamefont {Vaz}}, \bibinfo {author}
  {\bibfnamefont {N.~V.}\ \bibnamefont {Horne}}, \bibinfo {author}
  {\bibfnamefont {K.}~\bibnamefont {Bouzehouane}}, \bibinfo {author}
  {\bibfnamefont {K.}~\bibnamefont {Garcia}}, \bibinfo {author} {\bibfnamefont
  {C.}~\bibnamefont {Deranlot}}, \bibinfo {author} {\bibfnamefont
  {P.}~\bibnamefont {Warnicke}}, \bibinfo {author} {\bibfnamefont
  {P.}~\bibnamefont {Wohlh\"uter}}, \bibinfo {author} {\bibfnamefont {J.~M.}\
  \bibnamefont {George}}, \bibinfo {author} {\bibfnamefont {M.}~\bibnamefont
  {Weigand}}, \bibinfo {author} {\bibfnamefont {J.}~\bibnamefont {Raabe}},
  \bibinfo {author} {\bibfnamefont {V.}~\bibnamefont {Cros}}, \ and\ \bibinfo
  {author} {\bibfnamefont {A.}~\bibnamefont {Fert}},\ }\href@noop {} {\bibfield
   {journal} {\bibinfo  {journal} {Nature Nano.}\ }\textbf {\bibinfo {volume}
  {11}},\ \bibinfo {pages} {444} (\bibinfo {year} {2016})}\BibitemShut
  {NoStop}%
\bibitem [{\citenamefont {Feldtkeller}(1965)}]{feldtkeller1965a}%
  \BibitemOpen
  \bibfield  {author} {\bibinfo {author} {\bibfnamefont {E.}~\bibnamefont
  {Feldtkeller}},\ }\href@noop {} {\bibfield  {journal} {\bibinfo  {journal}
  {Z. Angew. Phys.}\ }\textbf {\bibinfo {volume} {19}},\ \bibinfo {pages} {530}
  (\bibinfo {year} {1965})}\BibitemShut {NoStop}%
\bibitem [{\citenamefont {Feldtkeller}(2017)}]{feldtkeller1965b}%
  \BibitemOpen
  \bibfield  {author} {\bibinfo {author} {\bibfnamefont {E.}~\bibnamefont
  {Feldtkeller}},\ }\href@noop {} {\bibfield  {journal} {\bibinfo  {journal}
  {IEEE Trans. Mag.}\ }\textbf {\bibinfo {volume} {53}},\ \bibinfo {pages}
  {0700308} (\bibinfo {year} {2017})}\BibitemShut {NoStop}%
\bibitem [{\citenamefont {D\"oring}(1968)}]{doering1968}%
  \BibitemOpen
  \bibfield  {author} {\bibinfo {author} {\bibfnamefont {W.}~\bibnamefont
  {D\"oring}},\ }\href@noop {} {\bibfield  {journal} {\bibinfo  {journal} {J.
  Appl. Phys.}\ }\textbf {\bibinfo {volume} {39}},\ \bibinfo {pages} {1006}
  (\bibinfo {year} {1968})}\BibitemShut {NoStop}%
\bibitem [{\citenamefont {Polyakov}(1974)}]{polyakov1974}%
  \BibitemOpen
  \bibfield  {author} {\bibinfo {author} {\bibfnamefont {A.~M.}\ \bibnamefont
  {Polyakov}},\ }\href@noop {} {\bibfield  {journal} {\bibinfo  {journal} {JETP
  Lett.}\ }\textbf {\bibinfo {volume} {20}},\ \bibinfo {pages} {430} (\bibinfo
  {year} {1974})}\BibitemShut {NoStop}%
\bibitem [{\citenamefont {Arrott}\ \emph {et~al.}(1979)\citenamefont {Arrott},
  \citenamefont {Heinrich},\ and\ \citenamefont {Aharoni}}]{arrott1979}%
  \BibitemOpen
  \bibfield  {author} {\bibinfo {author} {\bibfnamefont {A.~S.}\ \bibnamefont
  {Arrott}}, \bibinfo {author} {\bibfnamefont {B.}~\bibnamefont {Heinrich}}, \
  and\ \bibinfo {author} {\bibfnamefont {A.}~\bibnamefont {Aharoni}},\
  }\href@noop {} {\bibfield  {journal} {\bibinfo  {journal} {IEEE Trans.
  Magn.}\ }\textbf {\bibinfo {volume} {15}},\ \bibinfo {pages} {1228} (\bibinfo
  {year} {1979})}\BibitemShut {NoStop}%
\bibitem [{\citenamefont {Malozemoff}\ and\ \citenamefont
  {Slonczewski}(1979)}]{malozemoff1979}%
  \BibitemOpen
  \bibfield  {author} {\bibinfo {author} {\bibfnamefont {A.~P.}\ \bibnamefont
  {Malozemoff}}\ and\ \bibinfo {author} {\bibfnamefont {J.~C.}\ \bibnamefont
  {Slonczewski}},\ }\href@noop {} {\emph {\bibinfo {title} {Magnetic Domain
  Walls in Bubble Materials}}}\ (\bibinfo  {publisher} {Academic Press},\
  \bibinfo {year} {1979})\BibitemShut {NoStop}%
\bibitem [{\citenamefont {Volovik}(1987)}]{volovik1987}%
  \BibitemOpen
  \bibfield  {author} {\bibinfo {author} {\bibfnamefont {G.~E.}\ \bibnamefont
  {Volovik}},\ }\href@noop {} {\bibfield  {journal} {\bibinfo  {journal} {J.
  Phys. C}\ }\textbf {\bibinfo {volume} {20}},\ \bibinfo {pages} {L83}
  (\bibinfo {year} {1987})}\BibitemShut {NoStop}%
\bibitem [{\citenamefont {Kotiuga}(1989)}]{kotiuga1989}%
  \BibitemOpen
  \bibfield  {author} {\bibinfo {author} {\bibfnamefont {P.~R.}\ \bibnamefont
  {Kotiuga}},\ }\href@noop {} {\bibfield  {journal} {\bibinfo  {journal} {IEEE
  Trans. Mag.}\ }\textbf {\bibinfo {volume} {25}},\ \bibinfo {pages} {3476}
  (\bibinfo {year} {1989})}\BibitemShut {NoStop}%
\bibitem [{\citenamefont {Milde}\ \emph {et~al.}(2013)\citenamefont {Milde},
  \citenamefont {K\"ohler}, \citenamefont {Seidel}, \citenamefont {Eng},
  \citenamefont {Bauer}, \citenamefont {Chacon}, \citenamefont {Kindervater},
  \citenamefont {M\"uhlbauer}, \citenamefont {Pfleiderer}, \citenamefont
  {Buhrandt}, \citenamefont {Sch\"utte},\ and\ \citenamefont
  {Rosch}}]{milde2013}%
  \BibitemOpen
  \bibfield  {author} {\bibinfo {author} {\bibfnamefont {P.}~\bibnamefont
  {Milde}}, \bibinfo {author} {\bibfnamefont {D.}~\bibnamefont {K\"ohler}},
  \bibinfo {author} {\bibfnamefont {J.}~\bibnamefont {Seidel}}, \bibinfo
  {author} {\bibfnamefont {L.~M.}\ \bibnamefont {Eng}}, \bibinfo {author}
  {\bibfnamefont {A.}~\bibnamefont {Bauer}}, \bibinfo {author} {\bibfnamefont
  {A.}~\bibnamefont {Chacon}}, \bibinfo {author} {\bibfnamefont
  {J.}~\bibnamefont {Kindervater}}, \bibinfo {author} {\bibfnamefont
  {S.}~\bibnamefont {M\"uhlbauer}}, \bibinfo {author} {\bibfnamefont
  {C.}~\bibnamefont {Pfleiderer}}, \bibinfo {author} {\bibfnamefont
  {S.}~\bibnamefont {Buhrandt}}, \bibinfo {author} {\bibfnamefont
  {C.}~\bibnamefont {Sch\"utte}}, \ and\ \bibinfo {author} {\bibfnamefont
  {A.}~\bibnamefont {Rosch}},\ }\href@noop {} {\bibfield  {journal} {\bibinfo
  {journal} {Science}\ }\textbf {\bibinfo {volume} {340}},\ \bibinfo {pages}
  {1076} (\bibinfo {year} {2013})}\BibitemShut {NoStop}%
\bibitem [{\citenamefont {Donnelly}\ \emph {et~al.}(2017)\citenamefont
  {Donnelly}, \citenamefont {Guizar-Sicairos}, \citenamefont {Scagnoli},
  \citenamefont {Gliga}, \citenamefont {Holler}, \citenamefont {Raabe},\ and\
  \citenamefont {Heyderman}}]{heyderman2017}%
  \BibitemOpen
  \bibfield  {author} {\bibinfo {author} {\bibfnamefont {C.}~\bibnamefont
  {Donnelly}}, \bibinfo {author} {\bibfnamefont {M.}~\bibnamefont
  {Guizar-Sicairos}}, \bibinfo {author} {\bibfnamefont {V.}~\bibnamefont
  {Scagnoli}}, \bibinfo {author} {\bibfnamefont {S.}~\bibnamefont {Gliga}},
  \bibinfo {author} {\bibfnamefont {M.}~\bibnamefont {Holler}}, \bibinfo
  {author} {\bibfnamefont {J.}~\bibnamefont {Raabe}}, \ and\ \bibinfo {author}
  {\bibfnamefont {L.~J.}\ \bibnamefont {Heyderman}},\ }\href@noop {} {\bibfield
   {journal} {\bibinfo  {journal} {Nature}\ }\textbf {\bibinfo {volume}
  {547}},\ \bibinfo {pages} {328} (\bibinfo {year} {2017})}\BibitemShut
  {NoStop}%
\bibitem [{\citenamefont {Kanazawa}\ \emph {et~al.}(2017)\citenamefont
  {Kanazawa}, \citenamefont {White}, \citenamefont {Ronnow}, \citenamefont
  {Dewhurst}, \citenamefont {Morikawa}, \citenamefont {Shibata}, \citenamefont
  {Arima}, \citenamefont {Kagawa}, \citenamefont {Tsukazaki}, \citenamefont
  {Kozuka}, \citenamefont {Ichikawa}, \citenamefont {Kawasaki},\ and\
  \citenamefont {Tokura}}]{kanazawa2017}%
  \BibitemOpen
  \bibfield  {author} {\bibinfo {author} {\bibfnamefont {N.}~\bibnamefont
  {Kanazawa}}, \bibinfo {author} {\bibfnamefont {J.~S.}\ \bibnamefont {White}},
  \bibinfo {author} {\bibfnamefont {H.~M.}\ \bibnamefont {Ronnow}}, \bibinfo
  {author} {\bibfnamefont {C.~D.}\ \bibnamefont {Dewhurst}}, \bibinfo {author}
  {\bibfnamefont {D.}~\bibnamefont {Morikawa}}, \bibinfo {author}
  {\bibfnamefont {K.}~\bibnamefont {Shibata}}, \bibinfo {author} {\bibfnamefont
  {T.}~\bibnamefont {Arima}}, \bibinfo {author} {\bibfnamefont
  {F.}~\bibnamefont {Kagawa}}, \bibinfo {author} {\bibfnamefont
  {A.}~\bibnamefont {Tsukazaki}}, \bibinfo {author} {\bibfnamefont
  {Y.}~\bibnamefont {Kozuka}}, \bibinfo {author} {\bibfnamefont
  {M.}~\bibnamefont {Ichikawa}}, \bibinfo {author} {\bibfnamefont
  {M.}~\bibnamefont {Kawasaki}}, \ and\ \bibinfo {author} {\bibfnamefont
  {Y.}~\bibnamefont {Tokura}},\ }\href@noop {} {\bibfield  {journal} {\bibinfo
  {journal} {Phys. Rev. B}\ }\textbf {\bibinfo {volume} {96}},\ \bibinfo
  {pages} {220414(R)} (\bibinfo {year} {2017})}\BibitemShut {NoStop}%
\bibitem [{\citenamefont {Fujishiro}\ \emph {et~al.}(2019)\citenamefont
  {Fujishiro}, \citenamefont {Kanazawa}, \citenamefont {Nakajima},
  \citenamefont {Yu}, \citenamefont {Ohishi}, \citenamefont {Kawamura},
  \citenamefont {Kakurai}, \citenamefont {Arima}, \citenamefont {Mitamura},
  \citenamefont {Miyake}, \citenamefont {Akiba}, \citenamefont {Tokunaga},
  \citenamefont {Matsuo}, \citenamefont {Kindo}, \citenamefont {Koretsune},
  \citenamefont {Arita},\ and\ \citenamefont {Tokura}}]{tokura2019}%
  \BibitemOpen
  \bibfield  {author} {\bibinfo {author} {\bibfnamefont {Y.}~\bibnamefont
  {Fujishiro}}, \bibinfo {author} {\bibfnamefont {N.}~\bibnamefont {Kanazawa}},
  \bibinfo {author} {\bibfnamefont {T.}~\bibnamefont {Nakajima}}, \bibinfo
  {author} {\bibfnamefont {X.}~\bibnamefont {Yu}}, \bibinfo {author}
  {\bibfnamefont {K.}~\bibnamefont {Ohishi}}, \bibinfo {author} {\bibfnamefont
  {Y.}~\bibnamefont {Kawamura}}, \bibinfo {author} {\bibfnamefont
  {K.}~\bibnamefont {Kakurai}}, \bibinfo {author} {\bibfnamefont
  {T.}~\bibnamefont {Arima}}, \bibinfo {author} {\bibfnamefont
  {H.}~\bibnamefont {Mitamura}}, \bibinfo {author} {\bibfnamefont
  {A.}~\bibnamefont {Miyake}}, \bibinfo {author} {\bibfnamefont
  {K.}~\bibnamefont {Akiba}}, \bibinfo {author} {\bibfnamefont
  {M.}~\bibnamefont {Tokunaga}}, \bibinfo {author} {\bibfnamefont
  {A.}~\bibnamefont {Matsuo}}, \bibinfo {author} {\bibfnamefont
  {K.}~\bibnamefont {Kindo}}, \bibinfo {author} {\bibfnamefont
  {T.}~\bibnamefont {Koretsune}}, \bibinfo {author} {\bibfnamefont
  {R.}~\bibnamefont {Arita}}, \ and\ \bibinfo {author} {\bibfnamefont
  {Y.}~\bibnamefont {Tokura}},\ }\href@noop {} {\bibfield  {journal} {\bibinfo
  {journal} {Nature Comm.}\ }\textbf {\bibinfo {volume} {10}},\ \bibinfo
  {pages} {1059} (\bibinfo {year} {2019})}\BibitemShut {NoStop}%
\bibitem [{\citenamefont {Brown}(1963)}]{brown1963}%
  \BibitemOpen
  \bibfield  {author} {\bibinfo {author} {\bibfnamefont {W.~F.}\ \bibnamefont
  {Brown}},\ }\href@noop {} {\emph {\bibinfo {title} {Micromagnetics}}}\
  (\bibinfo  {publisher} {New York: Wiley},\ \bibinfo {year}
  {1963})\BibitemShut {NoStop}%
\bibitem [{\citenamefont {Sch\"utte}\ and\ \citenamefont
  {Rosch}(2014)}]{schuette2014}%
  \BibitemOpen
  \bibfield  {author} {\bibinfo {author} {\bibfnamefont {C.}~\bibnamefont
  {Sch\"utte}}\ and\ \bibinfo {author} {\bibfnamefont {A.}~\bibnamefont
  {Rosch}},\ }\href@noop {} {\bibfield  {journal} {\bibinfo  {journal} {Phys.
  Rev. B}\ }\textbf {\bibinfo {volume} {90}},\ \bibinfo {pages} {174432}
  (\bibinfo {year} {2014})}\BibitemShut {NoStop}%
\bibitem [{\citenamefont {Thiaville}\ \emph {et~al.}(2003)\citenamefont
  {Thiaville}, \citenamefont {Garc\'ia}, \citenamefont {Dittrich},
  \citenamefont {Miltat},\ and\ \citenamefont {Schrefl}}]{thiaville2003}%
  \BibitemOpen
  \bibfield  {author} {\bibinfo {author} {\bibfnamefont {A.}~\bibnamefont
  {Thiaville}}, \bibinfo {author} {\bibfnamefont {J.~M.}\ \bibnamefont
  {Garc\'ia}}, \bibinfo {author} {\bibfnamefont {R.}~\bibnamefont {Dittrich}},
  \bibinfo {author} {\bibfnamefont {J.}~\bibnamefont {Miltat}}, \ and\ \bibinfo
  {author} {\bibfnamefont {T.}~\bibnamefont {Schrefl}},\ }\href@noop {}
  {\bibfield  {journal} {\bibinfo  {journal} {Phys. Rev. B}\ }\textbf {\bibinfo
  {volume} {67}},\ \bibinfo {pages} {094410} (\bibinfo {year}
  {2003})}\BibitemShut {NoStop}%
\bibitem [{\citenamefont {Hertel}\ \emph {et~al.}(2007)\citenamefont {Hertel},
  \citenamefont {Gliga}, \citenamefont {F\"ahnle},\ and\ \citenamefont
  {Schneider}}]{hertel2007}%
  \BibitemOpen
  \bibfield  {author} {\bibinfo {author} {\bibfnamefont {R.}~\bibnamefont
  {Hertel}}, \bibinfo {author} {\bibfnamefont {S.}~\bibnamefont {Gliga}},
  \bibinfo {author} {\bibfnamefont {M.}~\bibnamefont {F\"ahnle}}, \ and\
  \bibinfo {author} {\bibfnamefont {C.~M.}\ \bibnamefont {Schneider}},\
  }\href@noop {} {\bibfield  {journal} {\bibinfo  {journal} {Phys. Rev. Lett.}\
  }\textbf {\bibinfo {volume} {98}},\ \bibinfo {pages} {117201} (\bibinfo
  {year} {2007})}\BibitemShut {NoStop}%
\bibitem [{\citenamefont {Braun}(1999)}]{braun1999}%
  \BibitemOpen
  \bibfield  {author} {\bibinfo {author} {\bibfnamefont {H.-B.}\ \bibnamefont
  {Braun}},\ }\href@noop {} {\bibfield  {journal} {\bibinfo  {journal} {J.
  Appl. Phys.}\ }\textbf {\bibinfo {volume} {85}},\ \bibinfo {pages} {6172}
  (\bibinfo {year} {1999})}\BibitemShut {NoStop}%
\bibitem [{\citenamefont {Col}\ \emph {et~al.}(2014)\citenamefont {Col},
  \citenamefont {Jamet}, \citenamefont {Rougemaille}, \citenamefont
  {Locatelli}, \citenamefont {Mentes}, \citenamefont {Burgos}, \citenamefont
  {Afid}, \citenamefont {Darques}, \citenamefont {Cagnon}, \citenamefont
  {Toussaint},\ and\ \citenamefont {Fruchart}}]{dacol2014}%
  \BibitemOpen
  \bibfield  {author} {\bibinfo {author} {\bibfnamefont {S.~D.}\ \bibnamefont
  {Col}}, \bibinfo {author} {\bibfnamefont {S.}~\bibnamefont {Jamet}}, \bibinfo
  {author} {\bibfnamefont {N.}~\bibnamefont {Rougemaille}}, \bibinfo {author}
  {\bibfnamefont {A.}~\bibnamefont {Locatelli}}, \bibinfo {author}
  {\bibfnamefont {T.~O.}\ \bibnamefont {Mentes}}, \bibinfo {author}
  {\bibfnamefont {B.~S.}\ \bibnamefont {Burgos}}, \bibinfo {author}
  {\bibfnamefont {R.}~\bibnamefont {Afid}}, \bibinfo {author} {\bibfnamefont
  {M.}~\bibnamefont {Darques}}, \bibinfo {author} {\bibfnamefont
  {L.}~\bibnamefont {Cagnon}}, \bibinfo {author} {\bibfnamefont {J.~C.}\
  \bibnamefont {Toussaint}}, \ and\ \bibinfo {author} {\bibfnamefont
  {O.}~\bibnamefont {Fruchart}},\ }\href@noop {} {\bibfield  {journal}
  {\bibinfo  {journal} {Phys. Rev. B}\ }\textbf {\bibinfo {volume} {89}},\
  \bibinfo {pages} {180405} (\bibinfo {year} {2014})}\BibitemShut {NoStop}%
\bibitem [{\citenamefont {Hertel}(2016)}]{hertel2016}%
  \BibitemOpen
  \bibfield  {author} {\bibinfo {author} {\bibfnamefont {R.}~\bibnamefont
  {Hertel}},\ }\href@noop {} {\bibfield  {journal} {\bibinfo  {journal} {J.
  Phys.: Condens. Matter}\ }\textbf {\bibinfo {volume} {28}},\ \bibinfo {pages}
  {483002} (\bibinfo {year} {2016})}\BibitemShut {NoStop}%
\bibitem [{\citenamefont {{M. Charilaou}}\ and\ \citenamefont
  {L\"offler}(2017)}]{charilaou2017}%
  \BibitemOpen
  \bibfield  {author} {\bibinfo {author} {\bibnamefont {{M. Charilaou}}}\ and\
  \bibinfo {author} {\bibfnamefont {J.~F.}\ \bibnamefont {L\"offler}},\
  }\href@noop {} {\bibfield  {journal} {\bibinfo  {journal} {Phys. Rev. B}\
  }\textbf {\bibinfo {volume} {024409}},\ \bibinfo {pages} {95} (\bibinfo
  {year} {2017})}\BibitemShut {NoStop}%
\bibitem [{\citenamefont {{M. Charilaou}}\ \emph {et~al.}(2018)\citenamefont
  {{M. Charilaou}}, \citenamefont {Braun},\ and\ \citenamefont
  {L\"offler}}]{charilaou2018}%
  \BibitemOpen
  \bibfield  {author} {\bibinfo {author} {\bibnamefont {{M. Charilaou}}},
  \bibinfo {author} {\bibfnamefont {H.-B.}\ \bibnamefont {Braun}}, \ and\
  \bibinfo {author} {\bibfnamefont {J.~F.}\ \bibnamefont {L\"offler}},\
  }\href@noop {} {\bibfield  {journal} {\bibinfo  {journal} {Phys. Rev. Lett.}\
  }\textbf {\bibinfo {volume} {121}},\ \bibinfo {pages} {097202} (\bibinfo
  {year} {2018})}\BibitemShut {NoStop}%
\bibitem [{\citenamefont {Im}\ \emph {et~al.}(2019)\citenamefont {Im},
  \citenamefont {Han}, \citenamefont {Jung}, \citenamefont {Yu}, \citenamefont
  {Lee}, \citenamefont {Yoon}, \citenamefont {Chao}, \citenamefont {Fischer},
  \citenamefont {Hong},\ and\ \citenamefont {Lee}}]{im2019}%
  \BibitemOpen
  \bibfield  {author} {\bibinfo {author} {\bibfnamefont {M.-Y.}\ \bibnamefont
  {Im}}, \bibinfo {author} {\bibfnamefont {H.-S.}\ \bibnamefont {Han}},
  \bibinfo {author} {\bibfnamefont {M.-S.}\ \bibnamefont {Jung}}, \bibinfo
  {author} {\bibfnamefont {Y.-S.}\ \bibnamefont {Yu}}, \bibinfo {author}
  {\bibfnamefont {S.}~\bibnamefont {Lee}}, \bibinfo {author} {\bibfnamefont
  {S.}~\bibnamefont {Yoon}}, \bibinfo {author} {\bibfnamefont {W.}~\bibnamefont
  {Chao}}, \bibinfo {author} {\bibfnamefont {P.}~\bibnamefont {Fischer}},
  \bibinfo {author} {\bibfnamefont {J.-I.}\ \bibnamefont {Hong}}, \ and\
  \bibinfo {author} {\bibfnamefont {K.-S.}\ \bibnamefont {Lee}},\ }\href@noop
  {} {\bibfield  {journal} {\bibinfo  {journal} {Nature Comm.}\ }\textbf
  {\bibinfo {volume} {10}},\ \bibinfo {pages} {593} (\bibinfo {year}
  {2019})}\BibitemShut {NoStop}%
\bibitem [{\citenamefont {Beg}\ \emph {et~al.}(2015{\natexlab{a}})\citenamefont
  {Beg}, \citenamefont {Carey}, \citenamefont {Wang}, \citenamefont
  {Cort\'es-Ortuno}, \citenamefont {Vousden}, \citenamefont {Bisotti},
  \citenamefont {Albert}, \citenamefont {Chernyshenko}, \citenamefont
  {Hovorka}, \citenamefont {Stamps},\ and\ \citenamefont {Fangohr}}]{beg2019}%
  \BibitemOpen
  \bibfield  {author} {\bibinfo {author} {\bibfnamefont {M.}~\bibnamefont
  {Beg}}, \bibinfo {author} {\bibfnamefont {R.}~\bibnamefont {Carey}}, \bibinfo
  {author} {\bibfnamefont {W.}~\bibnamefont {Wang}}, \bibinfo {author}
  {\bibfnamefont {D.}~\bibnamefont {Cort\'es-Ortuno}}, \bibinfo {author}
  {\bibfnamefont {M.}~\bibnamefont {Vousden}}, \bibinfo {author} {\bibfnamefont
  {M.-A.}\ \bibnamefont {Bisotti}}, \bibinfo {author} {\bibfnamefont
  {M.}~\bibnamefont {Albert}}, \bibinfo {author} {\bibfnamefont
  {D.}~\bibnamefont {Chernyshenko}}, \bibinfo {author} {\bibfnamefont
  {O.}~\bibnamefont {Hovorka}}, \bibinfo {author} {\bibfnamefont {R.~L.}\
  \bibnamefont {Stamps}}, \ and\ \bibinfo {author} {\bibfnamefont
  {H.}~\bibnamefont {Fangohr}},\ }\href@noop {} {\bibfield  {journal} {\bibinfo
   {journal} {Sci. Rep.}\ }\textbf {\bibinfo {volume} {5}},\ \bibinfo {pages}
  {17137} (\bibinfo {year} {2015}{\natexlab{a}})}\BibitemShut {NoStop}%
\bibitem [{\citenamefont {Kanazawa}\ \emph {et~al.}(2016)\citenamefont
  {Kanazawa}, \citenamefont {Nii}, \citenamefont {Zhang}, \citenamefont
  {Mishchenki}, \citenamefont {Filippis}, \citenamefont {Kagawa}, \citenamefont
  {Iwasa}, \citenamefont {Nagaosa},\ and\ \citenamefont
  {Tokura}}]{kanazawa2016}%
  \BibitemOpen
  \bibfield  {author} {\bibinfo {author} {\bibfnamefont {N.}~\bibnamefont
  {Kanazawa}}, \bibinfo {author} {\bibfnamefont {Y.}~\bibnamefont {Nii}},
  \bibinfo {author} {\bibfnamefont {X.-X.}\ \bibnamefont {Zhang}}, \bibinfo
  {author} {\bibfnamefont {A.~S.}\ \bibnamefont {Mishchenki}}, \bibinfo
  {author} {\bibfnamefont {G.~D.}\ \bibnamefont {Filippis}}, \bibinfo {author}
  {\bibfnamefont {F.}~\bibnamefont {Kagawa}}, \bibinfo {author} {\bibfnamefont
  {Y.}~\bibnamefont {Iwasa}}, \bibinfo {author} {\bibfnamefont
  {N.}~\bibnamefont {Nagaosa}}, \ and\ \bibinfo {author} {\bibfnamefont
  {Y.}~\bibnamefont {Tokura}},\ }\href@noop {} {\bibfield  {journal} {\bibinfo
  {journal} {Nature Comm.}\ }\textbf {\bibinfo {volume} {7}},\ \bibinfo {pages}
  {11622} (\bibinfo {year} {2016})}\BibitemShut {NoStop}%
\bibitem [{\citenamefont {Wilhelm}\ \emph {et~al.}(2011)\citenamefont
  {Wilhelm}, \citenamefont {Baenitz}, \citenamefont {Schmidt}, \citenamefont
  {R\"o\ss{}ler}, \citenamefont {Leonov},\ and\ \citenamefont
  {Bogdanov}}]{wilhelm2011}%
  \BibitemOpen
  \bibfield  {author} {\bibinfo {author} {\bibfnamefont {H.}~\bibnamefont
  {Wilhelm}}, \bibinfo {author} {\bibfnamefont {M.}~\bibnamefont {Baenitz}},
  \bibinfo {author} {\bibfnamefont {M.}~\bibnamefont {Schmidt}}, \bibinfo
  {author} {\bibfnamefont {U.~K.}\ \bibnamefont {R\"o\ss{}ler}}, \bibinfo
  {author} {\bibfnamefont {A.~A.}\ \bibnamefont {Leonov}}, \ and\ \bibinfo
  {author} {\bibfnamefont {A.~N.}\ \bibnamefont {Bogdanov}},\ }\href@noop {}
  {\bibfield  {journal} {\bibinfo  {journal} {Phys. Rev. Lett.}\ }\textbf
  {\bibinfo {volume} {107}},\ \bibinfo {pages} {127203} (\bibinfo {year}
  {2011})}\BibitemShut {NoStop}%
\bibitem [{\citenamefont {Vansteenkiste}\ \emph {et~al.}(2014)\citenamefont
  {Vansteenkiste}, \citenamefont {Leliaert}, \citenamefont {Dvornik},
  \citenamefont {Helsen}, \citenamefont {Garcia-Sanchez},\ and\ \citenamefont
  {Waeyenberge}}]{mumax3}%
  \BibitemOpen
  \bibfield  {author} {\bibinfo {author} {\bibfnamefont {A.}~\bibnamefont
  {Vansteenkiste}}, \bibinfo {author} {\bibfnamefont {J.}~\bibnamefont
  {Leliaert}}, \bibinfo {author} {\bibfnamefont {M.}~\bibnamefont {Dvornik}},
  \bibinfo {author} {\bibfnamefont {M.}~\bibnamefont {Helsen}}, \bibinfo
  {author} {\bibfnamefont {F.}~\bibnamefont {Garcia-Sanchez}}, \ and\ \bibinfo
  {author} {\bibfnamefont {B.~V.}\ \bibnamefont {Waeyenberge}},\ }\href@noop {}
  {\bibfield  {journal} {\bibinfo  {journal} {AIP Adv.}\ }\textbf {\bibinfo
  {volume} {4}},\ \bibinfo {pages} {107133} (\bibinfo {year}
  {2014})}\BibitemShut {NoStop}%
\bibitem [{\citenamefont {Ericsson}\ \emph {et~al.}(1981)\citenamefont
  {Ericsson}, \citenamefont {Karner}, \citenamefont {H\"aggstr\"om},\ and\
  \citenamefont {Chandra}}]{ericsson1981}%
  \BibitemOpen
  \bibfield  {author} {\bibinfo {author} {\bibfnamefont {T.}~\bibnamefont
  {Ericsson}}, \bibinfo {author} {\bibfnamefont {W.}~\bibnamefont {Karner}},
  \bibinfo {author} {\bibfnamefont {L.}~\bibnamefont {H\"aggstr\"om}}, \ and\
  \bibinfo {author} {\bibfnamefont {K.}~\bibnamefont {Chandra}},\ }\href@noop
  {} {\bibfield  {journal} {\bibinfo  {journal} {Physica Scripta}\ }\textbf
  {\bibinfo {volume} {23}},\ \bibinfo {pages} {1118} (\bibinfo {year}
  {1981})}\BibitemShut {NoStop}%
\bibitem [{\citenamefont {Yamadaa}\ \emph {et~al.}(2003)\citenamefont
  {Yamadaa}, \citenamefont {Teraoa}, \citenamefont {Ohtab},\ and\ \citenamefont
  {Kulatov}}]{yamada2003}%
  \BibitemOpen
  \bibfield  {author} {\bibinfo {author} {\bibfnamefont {H.}~\bibnamefont
  {Yamadaa}}, \bibinfo {author} {\bibfnamefont {K.}~\bibnamefont {Teraoa}},
  \bibinfo {author} {\bibfnamefont {H.}~\bibnamefont {Ohtab}}, \ and\ \bibinfo
  {author} {\bibfnamefont {E.}~\bibnamefont {Kulatov}},\ }\href@noop {}
  {\bibfield  {journal} {\bibinfo  {journal} {Physica B}\ }\textbf {\bibinfo
  {volume} {329}},\ \bibinfo {pages} {1131} (\bibinfo {year}
  {2003})}\BibitemShut {NoStop}%
\bibitem [{\citenamefont {Beg}\ \emph {et~al.}(2015{\natexlab{b}})\citenamefont
  {Beg}, \citenamefont {Carey}, \citenamefont {Wang}, \citenamefont
  {Cort\'es-Ortuno}, \citenamefont {Vousden}, \citenamefont {Bisotti},
  \citenamefont {Albert}, \citenamefont {Chernyshenko}, \citenamefont
  {Hovorka}, \citenamefont {Stamps},\ and\ \citenamefont {Fangohr}}]{beg2015}%
  \BibitemOpen
  \bibfield  {author} {\bibinfo {author} {\bibfnamefont {M.}~\bibnamefont
  {Beg}}, \bibinfo {author} {\bibfnamefont {R.}~\bibnamefont {Carey}}, \bibinfo
  {author} {\bibfnamefont {W.}~\bibnamefont {Wang}}, \bibinfo {author}
  {\bibfnamefont {D.}~\bibnamefont {Cort\'es-Ortuno}}, \bibinfo {author}
  {\bibfnamefont {M.}~\bibnamefont {Vousden}}, \bibinfo {author} {\bibfnamefont
  {M.-A.}\ \bibnamefont {Bisotti}}, \bibinfo {author} {\bibfnamefont
  {M.}~\bibnamefont {Albert}}, \bibinfo {author} {\bibfnamefont
  {D.}~\bibnamefont {Chernyshenko}}, \bibinfo {author} {\bibfnamefont
  {O.}~\bibnamefont {Hovorka}}, \bibinfo {author} {\bibfnamefont {R.~L.}\
  \bibnamefont {Stamps}}, \ and\ \bibinfo {author} {\bibfnamefont
  {H.}~\bibnamefont {Fangohr}},\ }\href@noop {} {\bibfield  {journal} {\bibinfo
   {journal} {Sci. Rep.}\ }\textbf {\bibinfo {volume} {5}},\ \bibinfo {pages}
  {17137} (\bibinfo {year} {2015}{\natexlab{b}})}\BibitemShut {NoStop}%
\bibitem [{\citenamefont {Andreas}\ \emph {et~al.}(2014)\citenamefont
  {Andreas}, \citenamefont {K\'akay},\ and\ \citenamefont
  {Hertel}}]{andreas2014}%
  \BibitemOpen
  \bibfield  {author} {\bibinfo {author} {\bibfnamefont {C.}~\bibnamefont
  {Andreas}}, \bibinfo {author} {\bibfnamefont {A.}~\bibnamefont {K\'akay}}, \
  and\ \bibinfo {author} {\bibfnamefont {R.}~\bibnamefont {Hertel}},\
  }\href@noop {} {\bibfield  {journal} {\bibinfo  {journal} {Phys. Rev. B}\
  }\textbf {\bibinfo {volume} {89}},\ \bibinfo {pages} {134403} (\bibinfo
  {year} {2014})}\BibitemShut {NoStop}%
\bibitem [{\citenamefont {Hubert}\ and\ \citenamefont
  {Sch\"afer}(1998)}]{hubert1998}%
  \BibitemOpen
  \bibfield  {author} {\bibinfo {author} {\bibfnamefont {A.}~\bibnamefont
  {Hubert}}\ and\ \bibinfo {author} {\bibfnamefont {R.}~\bibnamefont
  {Sch\"afer}},\ }\href@noop {} {\emph {\bibinfo {title} {Magnetic Domains: The
  analysis of magnetic microstructures}}}\ (\bibinfo  {publisher} {Springer
  Berlin},\ \bibinfo {year} {1998})\BibitemShut {NoStop}%
\bibitem [{\citenamefont {Li}\ \emph {et~al.}()\citenamefont {Li},
  \citenamefont {Pierobon}, \citenamefont {Charilaou}, \citenamefont {Braun},
  \citenamefont {Walet}, \citenamefont {L\"offler}, \citenamefont {Miles},\
  and\ \citenamefont {Moutafis}}]{yuli2019}%
  \BibitemOpen
  \bibfield  {author} {\bibinfo {author} {\bibfnamefont {Y.}~\bibnamefont
  {Li}}, \bibinfo {author} {\bibfnamefont {L.}~\bibnamefont {Pierobon}},
  \bibinfo {author} {\bibfnamefont {M.}~\bibnamefont {Charilaou}}, \bibinfo
  {author} {\bibfnamefont {H.-B.}\ \bibnamefont {Braun}}, \bibinfo {author}
  {\bibfnamefont {N.~R.}\ \bibnamefont {Walet}}, \bibinfo {author}
  {\bibfnamefont {J.~F.}\ \bibnamefont {L\"offler}}, \bibinfo {author}
  {\bibfnamefont {J.~J.}\ \bibnamefont {Miles}}, \ and\ \bibinfo {author}
  {\bibfnamefont {C.}~\bibnamefont {Moutafis}},\ }\href@noop {} {\bibinfo
  {journal} {arXiv:1911.12781}\ }\BibitemShut {NoStop}%
\bibitem [{\citenamefont {Pylypovski}\ \emph {et~al.}(2012)\citenamefont
  {Pylypovski}, \citenamefont {Sheka},\ and\ \citenamefont
  {Gaididei}}]{pylypovski2012}%
  \BibitemOpen
\bibfield  {journal} {  }\bibfield  {author} {\bibinfo {author} {\bibfnamefont
  {O.~V.}\ \bibnamefont {Pylypovski}}, \bibinfo {author} {\bibfnamefont
  {D.~D.}\ \bibnamefont {Sheka}}, \ and\ \bibinfo {author} {\bibfnamefont
  {Y.}~\bibnamefont {Gaididei}},\ }\href@noop {} {\bibfield  {journal}
  {\bibinfo  {journal} {Phys. Rev. B}\ }\textbf {\bibinfo {volume} {85}},\
  \bibinfo {pages} {224401} (\bibinfo {year} {2012})}\BibitemShut {NoStop}%
\bibitem [{\citenamefont {Kittel}(1949)}]{kittel1949}%
  \BibitemOpen
  \bibfield  {author} {\bibinfo {author} {\bibfnamefont {C.}~\bibnamefont
  {Kittel}},\ }\href@noop {} {\bibfield  {journal} {\bibinfo  {journal} {Rev.
  Mod. Phys.}\ }\textbf {\bibinfo {volume} {21}},\ \bibinfo {pages} {541}
  (\bibinfo {year} {1949})}\BibitemShut {NoStop}%
\bibitem [{\citenamefont {Kim}\ and\ \citenamefont
  {Tchernyshyov}(2013)}]{kim2013}%
  \BibitemOpen
  \bibfield  {author} {\bibinfo {author} {\bibfnamefont {S.~K.}\ \bibnamefont
  {Kim}}\ and\ \bibinfo {author} {\bibfnamefont {O.}~\bibnamefont
  {Tchernyshyov}},\ }\href@noop {} {\bibfield  {journal} {\bibinfo  {journal}
  {Phys. Rev. B}\ }\textbf {\bibinfo {volume} {174402}},\ \bibinfo {pages} {88}
  (\bibinfo {year} {2013})}\BibitemShut {NoStop}%
\bibitem [{\citenamefont {M\"uller}\ \emph {et~al.}(2019)\citenamefont
  {M\"uller}, \citenamefont {Hoffmann}, \citenamefont {Disselkamp},
  \citenamefont {Sch\"urhoff}, \citenamefont {Mavros}, \citenamefont
  {Sallermann}, \citenamefont {Kiselev}, \citenamefont {J\'onsson},\ and\
  \citenamefont {Bl\"ugel}}]{spirit}%
  \BibitemOpen
  \bibfield  {author} {\bibinfo {author} {\bibfnamefont {G.~D.}\ \bibnamefont
  {M\"uller}}, \bibinfo {author} {\bibfnamefont {M.}~\bibnamefont {Hoffmann}},
  \bibinfo {author} {\bibfnamefont {C.}~\bibnamefont {Disselkamp}}, \bibinfo
  {author} {\bibfnamefont {D.}~\bibnamefont {Sch\"urhoff}}, \bibinfo {author}
  {\bibfnamefont {S.}~\bibnamefont {Mavros}}, \bibinfo {author} {\bibfnamefont
  {M.}~\bibnamefont {Sallermann}}, \bibinfo {author} {\bibfnamefont {N.~S.}\
  \bibnamefont {Kiselev}}, \bibinfo {author} {\bibfnamefont {H.}~\bibnamefont
  {J\'onsson}}, \ and\ \bibinfo {author} {\bibfnamefont {S.}~\bibnamefont
  {Bl\"ugel}},\ }\href@noop {} {\bibfield  {journal} {\bibinfo  {journal}
  {Phys. Rev. B.}\ }\textbf {\bibinfo {volume} {99}},\ \bibinfo {pages}
  {224414} (\bibinfo {year} {2019})}\BibitemShut {NoStop}%
\end{thebibliography}

%merlin.mbs apsrev4-1.bst 2010-07-25 4.21a (PWD, AO, DPC) hacked
%Control: key (0)
%Control: author (72) initials jnrlst
%Control: editor formatted (1) identically to author
%Control: production of article title (-1) disabled
%Control: page (0) single
%Control: year (1) truncated
%Control: production of eprint (0) enabled
%

\end{document}